\documentstyle[epsfig]{aipproc}
\begin{document}
   \hyphenation{brems-strah-lung
                }
\title{Observational constraints on annihilation sites in 
   1E\,1740.7--2942 \\ and Nova Muscae}
\author{Igor V.\ Moskalenko$^\star$ and Elisabeth Jourdain$^{\dagger}$}
\address{
$^\star$Max-Planck-Institut f\"ur extraterrestrische Physik,
       D-85740 Garching, Germany \\
$^{\dagger}$Centre d'Etude Spatiale des Rayonnements, 
      31028 Toulouse Cedex, France}
\maketitle

\begin{abstract}
The region of the Galactic center contains several sources which
demonstrate their activity at various wavelengths and particularly
above several hundred keV \cite{Churazov94}. Escape of positrons from
such a source or several sources into the interstellar medium, where
they slow down and annihilate, can account for the 511 keV narrow line
observed from this direction. 1E\,1740.7--2942 object has been proposed
as the most likely candidate to be responsible for this variable source
of positrons \cite{Ramaty92}.  Besides, Nova Muscae shows a spectrum
which is consistent with Comptonization by a thermal plasma
$kT_e\lesssim100$ keV in its hard X-ray part, while a relatively narrow
annihilation line observed by SIGMA on Jan.\ 20--21, 1991 implies that
positrons annihilate in a much colder medium
\cite{Gilfanov91,Goldwurm92}.

We estimate the electron number density and the size of the emitting
regions suggesting that annihilation features observed by SIGMA from
Nova Muscae and 1E\,1740.7--2942 are due to the positron slowing down
and annihilation in thermal plasma. We show that in the case of Nova
Muscae the observed radiation is coming from a pair plasma stream
($n_{e^+}\approx n_{e^-}$) rather than from a gas cloud. We argue that
two models are probably relevant to the 1E source: annihilation in
(hydrogen) plasma $n_{e^+}\lesssim n_{e^-}$ at rest, and annihilation
in the pair plasma stream, which involves matter from the source
environment.

\end{abstract}

\section*{Observations}
\begin{figure}[t]
   \begin{picture}(139,70)(0,0)
      \put(2.5,0){\makebox(90,70)[tr]{%
         \psfig{file=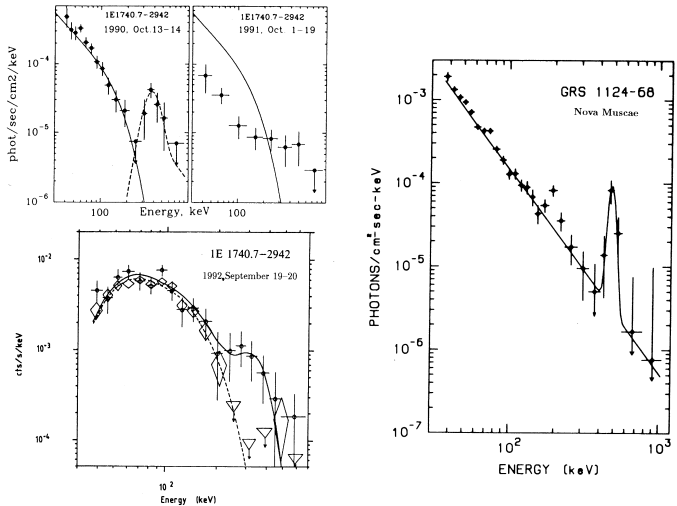,height=70mm,clip=}}}
      \put(91.8,0){ \begin{minipage}[b]{46.2mm}
         \caption[]{%
Energy spectra of 1E\,1740
\cite{Churazov94,Bouchet91,Sunyaev91,Cordier93,Churazov93} and NM
\cite{Goldwurm92} observed by SIGMA are shown together with fits of the
authors. For Sept.\ 1992 flare shown is counts s$^{-1}$ keV$^{-1}$.
The dashed line in the upper left panel shows the annihilation line
shape for Gaussian-like injection of energetic particles into the
thermal plasma of $kT=35$ keV for $E/A=20$.  The line is shifted left
to approach the data.}
         \label{fig1}
      \end{minipage}}
   \end{picture}
\end{figure}

Observations with the SIGMA telescope have revealed annihilation
features in the vicinity of $\sim500$ keV in spectra of two Galactic
black hole candidates (Fig.~\ref{fig1}): 1E\,1740.7--2942 (1E\,1740)
and Nova Muscae (NM).  Three times a broad excess in the 200--500 keV
region was observed in the 1E\,1740 emission spectra
\cite{Bouchet91,Sunyaev91,Cordier93,Churazov93}.  The features detected
on Oct.\ 13--14, 1990 and Sept.\ 19--20, 1992 showed similar fluxes and
line widths, the lifetime was restricted by 1--3 days. In Oct.\ 1991 an
excess at high energies was observed during 19 days and was not so
intensive as two others: the average flux was
$(1.9\pm0.6)\times10^{-3}$ phot cm$^{-2}$ s$^{-1}$ in the 300--600 keV
region.  On Jan.\ 20--21, 1991, the NM spectrum showed a clear emission
feature near 500 keV with the intrinsic line width $\lesssim58$ keV
\cite{Goldwurm92,Sunyaev92}.  Meanwhile, during all periods of
observation the hard X-ray emission, $\lesssim200$ keV, was found to be
consistent with the same law.  Observations of NM after the X-ray flare
are well fitted by a power-law of index $2.4-2.5$, the spectrum of
1E\,1740 is well described by Sunyaev-Titarchuk model \cite{ST80} with
$kT\approx35-60$ keV, $\tau\approx1.1-1.9$.  The observational data
\cite{Goldwurm92,Sunyaev92,Bouchet91,Sunyaev91,Cordier93} are
summarized in the first part of Table~\ref{tab1}.

\section*{Analysis and Discussion}
The spectral features observed by SIGMA are, commonly believed, related
to $e^+e^-$-annihilation. Relatively small line widths imply that the
temperature of the emitting region is quite low, $kT\approx35-45$ keV
for 1E\,1740 and $4-5$ keV for NM. Since the hard X-ray spectra showed
no changes, most probably that $e^+e^-$-pairs produced somewhere close
to the central object were injected into surrounding space where they
cool and annihilate.  Radiation pressure of a near-Eddington source
alone can accelerate $e^+e^-$-plasma up to the bulk Lorentz factor of
$\gamma_0\sim2-5$ \cite{Kovner84}, while Comptonization by the emergent
radiation field could provide a mechanism for cooling the plasma which
further annihilate `in flight' (for a discussion see
\cite{Gilfanov91}). If there is enough matter around a source, then
particles slow down due to Coulomb collisions and annihilate in the
medium. We explore further this last possibility by checking whether
the inferred parameters of the emitting region are consistent with
those obtained by other ways (for details see \cite{MJ97}).  We assume
single and short particle ejection on a timescale of hours; since the
ejection would probably impact on the whole spectrum, longer spectral
changes would be observable.

\begin{figure}[t] 
   \begin{picture}(139,37)(0,0)
      \put(-4,0){\makebox(45,42)[tl]{%
         \psfig{file=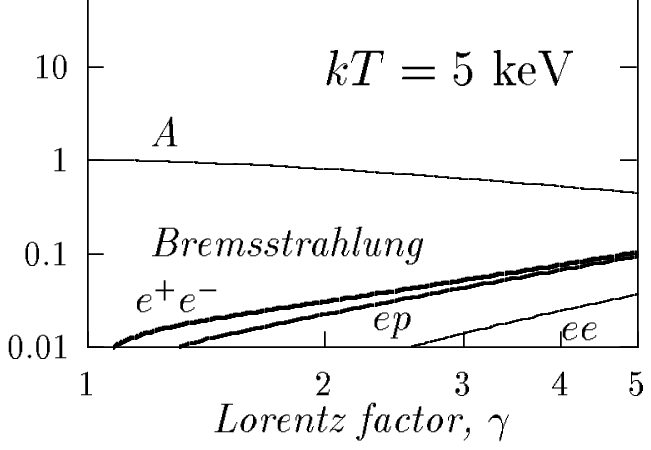,height=200mm,clip=}}}
      \put(44,0){\makebox(45,42)[tl]{%
         \psfig{file=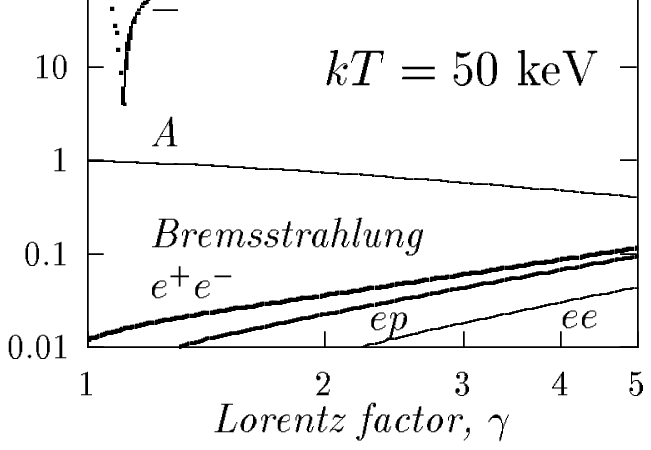,height=200mm,clip=}}}
      \put(97,0){ \begin{minipage}[b]{40mm}
         \caption[]{%
The dimensionless annihilation rate (A), and energy losses due to
brems\-strahlung and Coulomb scattering (C) in hydrogen plasmas
\cite{MJ97}.}
         \label{fig2}
      \end{minipage}}
   \end{picture}
\end{figure}

The relevant energy loss rates $|{d\gamma\over dt}|$ and annihilation
rate per one positron are shown in Fig.~\ref{fig2} in units $n_e\pi
r_e^2$.  Annihilation rate is small in comparison with the relaxation
rate, thus most of positrons annihilate after their distribution
approaches the steady-state one.

Suggesting that the energetic particles slow down due to Coulomb
scattering in the surrounding matter, one can estimate its (electron)
number density $n_-\approx{\gamma_0-1\over\pi r_e^2
c\,\Delta_i}|{d\gamma\over dt}|^{-1}$, where $\gamma_0$ is the initial
Lorentz factor of the plasma stream, $c$ is the light speed, and
$\Delta_i$ is the time scale of the annihilation line appearance.
Taking a reasonable value for the bulk Lorentz factor,
$\gamma_0\approx3$, one can obtain $n_-^{\sc 1e}\approx2.2\times10^7$
cm$^{-3}$ $(\Delta_i/ 2{\rm\ days})^{-1}$, and $n_-^{\sc
nm}\approx1.5\times10^8$ cm$^{-3}$ $(\Delta_i/5{\rm\ hr})^{-1}$.

If the particles were injected into the medium only once, then the
annihilation time scale is $\Delta_d\approx{1\over\pi r_e^2 c\,n_-A}$.
It yields one more estimate of the number density in the emitting
region $n_-\approx{1\over\pi r_e^2c\,\Delta_d\,A}\approx
1.55\times10^9{\rm\ cm}^{-3}\ (\Delta_d/1{\rm\ day})^{-1}$, where we
put $A=A(1)\approx1$ (Fig.~\ref{fig2}).  Total duration of this state
is $\Delta_d^{\sc 1e}\approx18-70$ hr, and $\Delta_d^{\sc nm}\le10$
days, that gives $n_-^{\sc 1e}\approx(5-20)\times10^8$ cm$^{-3}$ and
$n_-^{\sc nm}\approx1.5\times10^8$ cm$^{-3}$
$(\Delta_d/10{\rm\ days})^{-1}$, correspondingly. The values obtained
restrict the electron number density in the volume where particles slow
down and annihilate (Table~\ref{tab1}).

The time scales $\Delta_{i,d}$ are connected $\Delta_d/\Delta_i={1\over
A(\gamma_0-1)}|{d\gamma\over dt}|$, which is supported also by 1992
Sept.\ 19--20 observation. Therefore, the annihilation rise time on
1990 Oct.\ 13--14 should be $\Delta_i\approx1-2$ hr for consistency.

The size of the emitting region $\lambda$ can be estimated from a
relation $2n_+\lambda^3\approx\Delta_d L_{500}$ if we assume $n_+\le
n_-$ for the positron number density. It gives \linebreak $\lambda^{\sc
1e}\gtrsim1.34\times10^{13}$ cm $(\Delta_d/1{\rm\ day})^{2/3}
\approx(1.1-2.7)\times10^{13}$ cm and \linebreak $\lambda^{\sc
nm}\gtrsim1.3\times10^{13}$ cm $(\Delta_d/10{\rm\ days})^{2/3}$, while
the obvious upper limits are \linebreak $\lambda^{\sc 1e}<
c\Delta_i\approx2.2\times10^{14}$ cm $(\Delta_i/2{\rm\ hr})$ and
$\lambda^{\sc nm}\le5\times10^{14}$ cm.

\begin{table}[t!]
   \caption[]{\label{tab1}
Observational data and parameters of the emitting region.}
      \begin{tabular}{llll}
         \noalign{\vskip -0.5ex}
& \multicolumn{2}{c}{1E\,1740.7--2942} & \\
         \cline{2-3}
& \scriptsize 1990 Oct.13--14 & \scriptsize 1992 Sep.19--20 & 
                               \raisebox{2ex}[0pt]{\ Nova Muscae}\\
         \hline
         \noalign{\smallskip}
Annihilation rise time, $\Delta_i$
  & $\lesssim2$ days\tablenote{Our estimation is 1--2 hr;\hfill
  $^{\rm b}$1E\,1740: for 8.5 kpc distance; NM: for 1 kpc distance.}
                                          & few hours & $\sim5$ hr \\
Annihilation lifetime, $\Delta_d$
  & 18--70 hr        & 27--75 hr       & $\lesssim10$ days\\
Annihil.~line flux,\,$F_{500}$\,(phot cm$^{-2}$s$^{-1}$)
  & $10^{-2}$        & $4.3\times10^{-3}$ & $6\times10^{-3}$\\
Total line flux$^{\rm b}$, $L_{500}$ (photons s$^{-1}$)
  & $8.6\times10^{43}$ & $3.7\times10^{43}$ & $7.2\times10^{41}$ \\
Line width, $W$ (keV) & 240   & 180    & 40 \\
Column density, $N_{\sc h}$ (cm$^{-2}$)
   & \multicolumn{2}{c}{$\sim10^{23}$} & $\sim10^{21}$  \\
         \cline{1-1}
         \noalign{\smallskip}
Plasma temperature, $kT_e$ (keV)
   & \multicolumn{2}{c}{$35-45$}       & $3-4$ \\
Coulomb energy loss rate, $|d\gamma/dt|$
   & \multicolumn{2}{c}{70}            & 100   \\
Annihilation rate, $A$
   & \multicolumn{2}{c}{1}             & 1     \\
Electron number density, $n_-$ (cm$^{-3}$)
 &\multicolumn{2}{c}{$(5-20)\times10^8$} & $1.5\times10^8$ \\
Size of the emitting region, $\lambda$ (cm)
 & \multicolumn{2}{c}{$(1.1-20)\times10^{13}$}
 & $(1.3-50)\!\times\!10^{13}$\\
      \end{tabular}
\end{table}

The column density of the medium where particles slow down should
exceed the value $N_{\sc h}\sim\lambda n_-$, viz.\ $N_{\sc h} ^{\sc
nm}\gtrsim2\times10^{21}$ cm$^{-2}$ $(\Delta_d/ 10$ days$)^{-1/3}$, and
\linebreak $2.1\times10^{22}$ cm$^{-2}$ $(\Delta_d/ 1$ day$)^{-1/3}
\lesssim N_{\sc h}^{\sc 1e}< c\Delta_in_-\approx1.1\times10^{23}$
cm$^{-2}$.  The column density of the gas cloud measured along the line
of sight, where 1E\,1740 embedded, is high enough $N_{\sc
h}\approx3\times10^{23}$ cm$^{-2}$ \cite{BL91,Mirabel91}.  Note that
ASCA measurements give the column density {\it to this source} of
$\approx8\times10^{22}$ cm$^{-2}$ \cite{Sheth96}. For NM the
corresponding value is $N_{\sc h}\sim10^{21}$ cm$^{-2}$
\cite{Greiner91}, less or marginally close to the obtained lower
limit.  If, on contrary, one suggests $n_+\ll n_-$ it yields a
condition $N_{\sc h}\gg2\times10^{21}$ cm$^{-2}$ $(\Delta_d/
10{\rm\ days})^{-1/3}$, which considerably exceeds the measured value.

These estimations imply that the 500 keV emission observed from NM was
coming from $e^+e^-$-plasma jet ($n_+\approx n_-$) rather than from
particles injected into a gas cloud\footnote{A possibility that NM lies
in front of a large gas cloud can not be totally excluded. In this
case, particles could be injected into this cloud, away from the
observer.} ($n_+\ll n_-$), therefore, particles have to annihilate `in
flight' producing a relatively narrow line shifted dependently on the
jet orientation. If so, then our estimation of $n_-$ from annihilation
time scale gives the average electron/positron number density in the
jet, fixing the total volume as $\lambda^3\sim2\times10^{39}$ cm$^3$
$(\Delta_d/10{\rm\ days})^2$. The reported 6\%--7\% redshift
\cite{Goldwurm92,Sunyaev92} supports probably the annihilation-in-jet
hypothesis, although its statistical significance is small. The large
size of the emitting region and a small width of the line, both except
the gravitational origin of the redshift.

For the emitting region in 1E\,1740 our estimations give $n_-\gtrsim
n_+$.  Two events, Oct.\ 1990 and Sept.\ 1992, have shown similar
parameters, which are consistent with single particle injection into
the thermal (hydrogen) plasma.  The redshift of the line $\sim25$\%
\cite{Bouchet91,Sunyaev91,Cordier93} implies that positrons probably
annihilate in a plasma stream moving away from the observer, since the
size of the emitting region is too large and rules out its
gravitational nature.  A natural explanation of this controversial
picture is that the plasma stream captures matter from the source
environment and annihilation occurs in a moving plasma volume. The
value of $n_-$ obtained is then the average electron number density in
the jet, $\lambda^3\gtrsim2.4\times10^{39}$ cm$^3$
$(\Delta_d/1{\rm\ day})^2$ is its total volume, and the jet length
should be of the order of $\sim0.2c\Delta_d\approx5.2\times10^{14}$ cm
$(\Delta_d/1{\rm\ day})$.

While some part of the pair plasma annihilates near 1E\,1740 producing
the broad line, the remainder could escape into a molecular cloud,
which was found to be associated with this source
\cite{BL91,Mirabel91}.  Taking $\sim10^5$ cm$^{-3}$ for the average
number density of the cloud one can obtain for the slowing down time
scale\footnote{ The annihilation lifetime $\Delta_d$ was obtained for
thermal plasma and is not valid for the cold medium where positrons
mostly annihilate in the bound (positronium) state.}
$\Delta_i\lesssim1$ year, the same as was obtained in \cite{Ramaty92}.
The size of the turbulent region caused by propagation of a dense jet
should also be of the order of 1 ly. It agrees well with the length
2--4 ly (15--30 arcsec at 8.5 kpc) of a double-sided radio jet
symmetrical about 1E\,1740 \cite{Mirabel92}.

If the lines from 1E\,1740 (Fig.~\ref{fig1}) were produced by
continuous injection of energetic particles, then observations of the
narrow 511 keV line emission from the Galactic center allows to put an
upper limit on the particle escape rate into the interstellar medium.
Taking $\tau_0=1$ year for the positron lifetime in $10^5$ cm$^{-3}$
dense molecular cloud \cite{Ramaty92}, and suggesting one hard state of
$\Delta_d\gtrsim2$ days long per period $\tau_0$, one can obtain an
escape rate $E/A\approx{F_{511}\,\tau_0\over
F_{500}\,\Delta_d}\lesssim20$, where we took $F_{511}\approx10^{-3}$
phot cm$^{-2}$ s$^{-1}$ for the narrow line intensity \cite{MLW94}, and
$F_{500}=10^{-2}$ phot cm$^{-2}$ s$^{-1}$ (Table 1). This is consistent
with 1990 Oct.\ 13--14 and 1992 Sept.\ 19--20 spectra; the dashed line
in Fig.~\ref{fig1} shows the annihilation line shape for Gaussian-like
particle injection, $\sim\exp[-(\gamma-4)^2]$, into the thermal plasma
of $kT=35$ keV for $E/A=20$ \cite{MJ97}. The longest hard state (19
days, Oct.\ 1991) with the average flux of
$F_{500}\approx2\times10^{-3}$ phot cm$^{-2}$ s$^{-1}$ places the upper
limit at almost the same level $E/A\approx10$.

\footnotesize
I.M.\ is grateful to the SIGMA team of CESR for hospitality and facilities.


\begin{references}
\bibitem{Churazov94} Churazov E., et al., {\it ApJS} {\bf 92}, 381 (1994)
\bibitem{Ramaty92} Ramaty R., et al., {\it ApJ} {\bf 392}, L63 (1992)
\bibitem{Gilfanov91} Gilfanov M., et al., {\it Soviet Astron.\
   Lett.}\ {\bf 17}, 437 (1991)
\bibitem{Goldwurm92} Goldwurm A., et al., {\it ApJ} {\bf 389}, L79 (1992)
\bibitem{Sunyaev92} Sunyaev R.\ A., et al., {\it ApJ} {\bf 389}, L75 (1992)
\bibitem{Bouchet91} Bouchet L., et al., {\it ApJ} {\bf 383}, L45 (1991)
\bibitem{Sunyaev91} Sunyaev R.\ A., et al., {\it ApJ} {\bf 383}, L49 (1991)
\bibitem{Cordier93} Cordier B., et al., {\it A\&A} {\bf 275}, L1 (1993)
\bibitem{Churazov93} Churazov E., et al., {\it ApJ} {\bf 407}, 752 (1993)
\bibitem{ST80} Sunyaev R.\ A., Titarchuk L.\ G., {\it A\&A} {\bf 86}, 121 (1980)
\bibitem{Kovner84} Kovner I., {\it A\&A} {\bf 141}, 341 (1984)
\bibitem{MJ97} Moskalenko I.\ V., Jourdain E., {\it A\&A } {\bf in press} (1997)
   -- astro-ph/9702071
\bibitem{BL91} Bally J., Leventhal M., {\it Nature} {\bf 353}, 234 (1991)
\bibitem{Mirabel91} Mirabel I.\ F., et al., {\it A\&A} {\bf 251}, L43 (1991)
\bibitem{Sheth96} Sheth S., et al., {\it ApJ} {\bf 468}, 755 (1996)
\bibitem{Greiner91} Greiner J., et al., in {\it Proc.\ Workshop on Nova
   Muscae}, ed.\ S.\ Brandt, Lyngby: Danish Space Research Inst., 1991, p.79
\bibitem{Mirabel92} Mirabel I.\ F., et al., {\it Nature} {\bf 358}, 215 (1992)
\bibitem{MLW94} Mahoney W.\ A., Ling J.\ C., Wheaton Wm.\ A., 
   {\it ApJS} {\bf 92}, 387 (1994)

\end{references}
\end{document}